\documentclass[11pt]{article}

\usepackage{a4}
\usepackage[fleqn]{amsmath}
\usepackage{hyperref}

\usepackage{fourier} 
\usepackage[scaled=0.875]{helvet} 

\RequirePackage{amsmath,amssymb,amsxtra,amsthm}

\RequirePackage{mathrsfs}

\RequirePackage{setspace}
\RequirePackage{array}
\RequirePackage{booktabs}

\RequirePackage{braket}

\RequirePackage[pdftex,final]{graphicx}
\graphicspath{{Plots/}}

\RequirePackage{colortbl}
\RequirePackage{xcolor}
\colorlet{titlerowcolor}{gray!15}

\usepackage{cite}

\numberwithin{equation}{section}
\numberwithin{table}{section}
\numberwithin{figure}{section}

\author{
  \begin{minipage}{1.00\linewidth}
    \vspace{1cm}
    \begin{center}
      \begin{small}
        \ \ \textbf{Carlo Angelantonj} $^{1,2}$ and \textbf{Ignatios Antoniadis}$^{3,4}$
     \end{small}
    \end{center}
    \vspace{.3cm} \hspace{0.75cm}\begin{minipage}{.85\linewidth}
      {\it \begin{footnotesize}
          \begin{itemize}
         \item[${}^1$] Dipartimento di Fisica, Universit\`a di Torino, and INFN Sezione di Torino
          \\
            Via Pietro Giuria 1, 10125 Torino, Italy
         \item[${}^2$] Arnold-Regge Center, Via Pietro Giuria 1, 10125 Torino, Italy 
         \item[${}^3$] Laboratoire de Physique Th\'eorique et Hautes \'Energies - LPTHE\\
Sorbonne Universit\'e, CNRS, 4 Place Jussieu, 75005 Paris, France
          \item[${}^4$]  Albert Einstein Center for Fundamental Physics,
Institute for Theoretical Physics,\\
University of Bern,
Sidlerstrasse 5, CH-3012 Bern, Switzerland
\end{itemize}
        \end{footnotesize}}
    \end{minipage}
    \vspace{1cm}
  \end{minipage}
}

\date{}

\title{\vspace{3cm}
  \begin{huge}
    \textbf{The String Geometry Behind Topological Amplitudes} 
  \end{huge}
}

\begin{document}

\begin{titlepage}
  \maketitle
  \thispagestyle{empty}

  \vspace{-14cm}
  \begin{flushright}
{\bf    \today}
   \end{flushright}

  \vspace{11cm}

  \begin{center}
    \textsc{Abstract}\\
  \end{center}

It is shown that the generating function of $\mathscr{N}=2$ topological strings, in the heterotic weak coupling limit, is identified with the partition function of a six-dimensio\-nal Melvin background. This background, which corresponds to an exact CFT, realises in string theory the  six-dimensional $\varOmega$-background of Nekrasov, in the case of opposite deformation parameters $\epsilon_1=-\epsilon_2$, thus providing the known perturbative part of the Nekrasov partition function in the field theory limit. The analysis is performed on both heterotic and type I strings and for the cases of ordinary $\mathscr{N}=2$ and $\mathscr{N}=2^*$ theories. 
 
\vfill

{\small
\begin{itemize}
\item[E-mail:] {\tt carlo.angelantonj@unito.it}
\\
{\tt ignatios.antoniadis@upmc.fr}
\end{itemize}
}
\vfill

\end{titlepage}

\setstretch{1.1}

\tableofcontents


\section{Introduction}

It is known~\cite{Antoniadis:1993ze} that a series of higher derivative $F$-terms of $\mathscr{N}=2$ supersymmetric compactifications of string theory in four dimensions, of the form $F_gW^{2g}$ with $W$ the Weyl superfield and $F_g$ a function of the vector moduli, is computed by the genus-$g$ partition function of a topological string~\cite{Bershadsky:1993cx} obtained by an appropriate twist~\cite{Witten:1988xj,Witten:1989ig} of the corresponding $\mathscr{N}=2$ superconformal $\sigma$-model describing the compactification on a six-dimensional Calabi-Yau (CY) manifold. An important property of $F_g$'s is the holomorphic anomaly expressed as a recursive differential equation that can be understood either from boundary contributions in the degeneration limit of Riemann surfaces within the topological theory~\cite{Bershadsky:1993cx}, or from non-local terms in the string effective action due to the propagation of massless states~\cite{Antoniadis:1993ze}. As a consequence of the heterotic-type II string duality, the $F_g$'s can be easily studied on the heterotic (or type I) side at the one loop level,  upon identifying the heterotic string dilaton with a particular $\mathscr{N}=2$ vector multiplet, corresponding to the base modulus of the CY manifolds which are $K3$ fibrations~\cite{Antoniadis:1995zn}. This has also the advantage of allowing a straightforward field theory limit which corresponds to a perturbative non-abelian gauge group enhancement, freezing the dilaton to a constant vacuum expectation value.

An interesting question is to understand the possible geometry which is generated by the topological amplitudes. The answer on the field theory side is given by the four-dimensional $\varOmega$ background, in the case of opposite deformation parameters $\epsilon_1=-\epsilon_2$~\cite{Moore:1997dj, Losev:1997wp, Nekrasov:2002qd, Nekrasov:2003rj}. From the point of view of the heterotic amplitudes, this condition on the $\epsilon$'s translates into a constant anti-self-dual field strength for the graviphoton $T$, which is the lowest component of the (chiral)  Weyl superfield $W=T+\theta R\theta+\dots$, with $R$ the anti-self-dual Riemann tensor ~\cite{deRoo:1980mm,Bergshoeff:1980is}. Notice that the topological string coupling is identified with $T$ (for $g\ge 1$). 

In this work, we address this question directly in string theory, both on the heterotic and type I sides, compactified on $K3\times T^2$, and  show that the emerging four-dimensional geometry is the so-called Melvin space, where string dynamics can be studied exactly~\cite{Russo:1995tj,Russo:1994cv}. This result agrees with and extends to a full string setup  previous studies  \cite{Hellerman:2011mv, Hellerman:2012zf, Orlando:2013yea, Lambert:2014fma}, where the classical low-energy dynamics of D-branes on Melvin spaces was shown to reproduce the equivariant action of \cite{Nekrasov:2003rj}.

The string dynamics on Melvin spaces is particularly simple since the geometry is flat, and in fact it can be realised as a freely acting orbifold where rotations on two planes are accompanied by a winding around a one-cycle of the $T^2$.  If the rotations are chosen to act equally in opposite directions, the background breaks half of the original supersymmetries, and thus preserves four supercharges.  We then compute the one loop string partition function, both for the heterotic and type I string, and show that it faithfully reproduces  the  corresponding $\mathscr{N}=2$ topological amplitudes.  We then conclude that the collective effect of the scattering of the $2g-2$ graviphotons back-reacts on the four-dimensional Minkowski space and generates the non-trivial Melvin geometry. Clearly, our string computation of the one-loop partition function reduces to the perturbative contribution of the Nekrasov free energy in the field theory limit.

The outline of the paper is the following. In Section 2, we give a brief overview of $\mathscr{N}=2$ topological amplitudes computed at one loop level on the heterotic side via a generating function, depending on the $T^2$ moduli and the topological string coupling $\lambda$. In Section 3, we extract from inspection of the generating function the underlying geometric background which turns out to be a four-dimensional (Eulcidean) Melvin space with non-trivial gauge fields and dilaton profile. Since  the graviphoton vertex involved in the topological amplitudes has a leg on the internal $T^2$, we extend appropriately the Melvin background to a six-dimensional space containing also the $T^2$. In Section 4, we analyse the quantum string propagation on generic Melvin spaces and we show that the problem is reduced to a freely acting orbifold acting on one non-compact complex coordinate, where the twist is accompanied with a shift along an extra circle. In Section 5, we extend the analysis to the full six-dimensional Melvin background, involving  opposite rotations in two planes, and combine it with the compactification on $K3$ in the orbifold limit to yield $\mathscr{N}=2$ supersymmetry. In Section 6, we use the above results to compute the partition function of the Heterotic string on  $(\text{Melvin}\otimes T^2)\times K3$ and we show that it coincides with the generating function of $\mathscr{N}=2$ and $\mathscr{N}=2^*$ topological amplitudes. In the following two sections, we extend the analysis to open strings. We quantise the string coordinates in Section 7,  and then in Section 8 we address the full ten-dimensional D-brane dynamics on ($\text{Melvin}\otimes T^2)\times K3$ both for the $\mathscr{N}=2$ and $\mathscr{N}=2^*$ gauge theories. In Section 9 we draw our conclusions and compare with the existing literature. The paper also comprises two appendices containing the definitions and main properties of the Dedekind-eta function and Jacobi-theta functions (Appendix A), as well as some useful properties of the 2d Narain lattice (Appendix B).

\section{Glimpses on Heterotic Topological Amplitudes}  \label{TopAmp}

The topological amplitudes corresponding to the $F^{g}\, W^{2g}$ higher derivative couplings with $\mathscr{N}=2$ supersymmetry can be computed in the heterotic string \cite{Antoniadis:1995zn} via the correlator
\begin{equation}
\bigg\langle V_h (p_1) \, V_h (\bar p_2) \, \prod_{i=1}^{g-1} V_F (p_1^{(i)})\, V_F (\bar p_2^{(i)}) \bigg\rangle
\end{equation}
where 
\begin{equation}
\begin{split}
V_h (p_1) &= \left(\partial Z^2 - i p_1 \chi^1 \chi^2 \right) \partial Z^2 \, e^{i p_1 Z^1}\,,
\\
V_h (\bar p_2) &= \left(\partial \bar Z^1 - i \bar p_2 \bar\chi^2 \bar\chi^1 \right) \bar\partial \bar Z^1 \, e^{i \bar p_2 \bar Z^2}\,,
\end{split}
\end{equation}
and
\begin{equation}
\begin{split}
V_F (p_1 ) &= \left(\partial X - i p_1 \chi^1 \varPsi \right) \, \bar\partial Z^2 \, e^{i p_1 Z^1 }\,,
\\
V_F (\bar p_2 ) &= \left( \partial X - i \bar p_2 \bar \chi^2 \varPsi \right) \bar\partial\bar Z^1 \, e^{i \bar p_2 \bar Z^2}\,,
\end{split}
\end{equation}
are the vertex operators for the anti-self-dual part of the Riemann tensor and for the graviphotons, respectively, in the zero ghost picture. 
The complex combinations
\begin{equation}
\begin{split}
Z^1 &= \frac{ X^1 - i X^2 }{\sqrt{2}}\,,
\\
Z^2 &= \frac{X^0 - i X^3 }{\sqrt{2}}\,,
\end{split}
\qquad
\begin{split}
\chi^1 &= \frac{\psi^1 - i \psi^2 }{\sqrt{2}}\,,
\\
\chi^2 &= \frac{\psi^0 - i \psi^3}{\sqrt{2}}\,,
\end{split}
\end{equation}
denote the bosonic and fermionic coordinates on the four-dimensional Minkowski space, while $X$ and its supersymmetric partner $\varPsi$ refer to a complex compact internal coordinate. The expectation value is evaluated in a heterotic background with $\mathscr{N}=2$ supersymmetry, which we assume to be described by the orbifold limit of $K3$. 

The choice of the space-time momenta, $p_1 \not=0, \ p_2 = \bar p_1 = \bar p_2 =0$ for the first set of vertex operators and $p_1 = p_2 = \bar p_1 = 0\, \  \bar p_2 \not=0$ for the second set, helps to simplify the computation of the amplitude, and yields the expression
\begin{equation}
\begin{split}
F_g &= - \frac{(4 \pi i )^{g-1}}{4\pi^2} \, \frac{1}{(g!)^2} \, \int_\mathscr{F} \frac{d^2\tau}{\tau_2^3}\, F (\bar \tau ) \, \bigg\langle \prod_{i=1}^g \int d^2 x_i \, Z^1 \bar\partial Z^2 (x_i) \, \prod_{j=1}^g \int d^2 y_j \, \bar Z^2 \bar \partial \bar Z^1 (y_j ) \bigg\rangle
\\
&\quad \times \sum_{m,n} \left( \frac{p_L}{2 T_2} \right)^{2g-2} \, q^{\frac{1}{4} |p_L|^2} \, \bar q ^{\frac{1}{4} |p_R|^2} \,,
\end{split}\label{Fg1}
\end{equation}
where
\begin{equation}
\begin{split}
p_L &=\frac{1}{\sqrt{T_2 U_2}} \left( m_2 - Um_1 +\bar T (n_1 + U n_2) \right) \,,
\\
p_R &=\frac{1}{\sqrt{T_2 U_2}} \left( m_2 - Um_1 + T (n_1 + U n_2) \right)\,,
\end{split}
\end{equation}
are the left-moving and right-moving momenta along the compact $T^2$ with complex structure $U=U_1 + i U_2$ and K\"ahler form $T=T_1+iT_2$. In computing the amplitude, one should sum over all spin structures with suitable GSO phases. However, one can show that even and odd spin structures give exactly the same contribution. As a result,
\begin{equation}
F (\bar \tau ) = \text{tr} \, (-1)^F\, q^{L_0 - c/24}\, \bar q ^{\bar L_0 - \bar c/24}
\end{equation}
is an  anti-holomorphic function of $\bar\tau$, since the presence of the operator $(-1)^F$ ensures that the contributions from the excitations of the supersymmetric (left-moving) sector vanish identically. 

In order to compute the correlation functions in \eqref{Fg1}, it is convenient to introduce the generating function
\begin{equation}
G (\lambda ; \tau , \bar \tau ) = \sum_{g=1}^\infty \frac{1}{(g!)^2} \, \left( \frac{\lambda}{\tau_2}\right)^{2g}\, \bigg\langle \prod_{i=1}^g \int d^2 x_i \, Z^1 \bar\partial Z^2 (x_i) \, \prod_{j=1}^g \int d^2 y_j \, \bar Z^2 \bar \partial \bar Z^1 (y_j ) \bigg\rangle \,,
\end{equation}
which is modular invariant if one postulates the transformation $\lambda \to \lambda / (c\bar \tau + d)$,
\begin{equation}
G \left( \frac{ \lambda}{c\bar \tau + d}, \frac{a  \tau + b}{c \tau + d}, \frac{a \bar \tau + b}{c\bar \tau + d} \right) = G (\lambda , \tau , \bar\tau )\,.
\end{equation}
The advantage of having introduced the generating function $G$ is that it can be expressed as the normalised functional integral
\begin{equation}
G (\lambda , \tau , \bar\tau ) = \frac{ \int \prod_{i=1,2} \mathscr{D} Z^i \mathscr{D} \bar Z^i \, \exp \left( - S + \frac{\lambda}{\tau_2} \int d^2 x \left( Z^1 \bar\partial Z^2 + \bar Z^2 \bar \partial \bar Z^1 \right) \right)}{\int \prod_{i=1,2} \mathscr{D} Z^i \mathscr{D} \bar Z^i \, \exp ( - S )}
\end{equation}
over the four bosonic coordinates of the non-compact space time, with 
\begin{equation}
S = \frac{1}{\pi} \int d^2 x\, \sum_{i=1,2} \left(\partial Z^i \bar \partial \bar Z^i + \partial \bar Z^i \bar\partial Z^i \right)
\end{equation}
the free-field Polyakov action. One is thus left with Gaussian integrals over bosonic coordinates, which can be straightforwardly computed using $\zeta$-function regularisation to get
\begin{equation}
G (\lambda , \tau , \bar\tau ) = \left( \frac{2 \pi i \lambda \bar\eta^3}{\bar\vartheta_1 (\lambda |\bar \tau )}\right)^2 \, e^{-\pi \lambda^2/\tau_2}\,.
\label{genfunctG}
\end{equation}
The non-holomorphic contribution to $G$ is a consequence of the $\zeta$-function regularisation of the functional determinant, and its presente is at the origin of the holomorphic anomaly for the $F_g$'s, since 
\begin{equation}
\partial_\tau G (\lambda , \tau , \bar \tau ) = - \frac{i\pi\lambda^2}{2\tau_2^2} \, G (\lambda , \tau , \bar \tau )\,.
\end{equation}

Given the result \eqref{genfunctG}, one can Taylor expand  $F(\lambda )$ to extract the topological amplitudes $F_g$ \cite{Antoniadis:1995zn}, 
\begin{equation}
\begin{split}
F (\lambda) &=  \sum_{g=1}^\infty \lambda^{2g}\, F_g
\\
&= -\frac{1}{4\pi^2} \int_\mathscr{F} \frac{d^2\tau}{\tau_2} \, F (\bar \tau )\, \sum_{m,n} \left( \frac{2\pi i \lambda \bar\eta^3}{\bar\vartheta_1 (\tilde\lambda |\bar \tau )}\right)^2\, e^{- \pi \tilde\lambda^2 /\tau_2} \, q^{\frac{1}{4} |p_L|^2}\, \bar q ^{\frac{1}{4} |p_R|^2} \,,
\end{split}\label{topoHam}
\end{equation}
where $\tilde \lambda = \sqrt{4\pi i} \lambda p_L \tau_2/ (2 T_2)$. 

A useful, alternative representation of the generating function is obtained upon Poisson summation over the two-dimensional KK momenta. Using the expansion of $\vartheta_1$ and the properties of the two-dimensional lattice collected in appendixes \ref{appJacobi} and \ref{NarainlattB}, one can write
\begin{equation}
F (\lambda ) = -\frac{1}{4\pi^2} \int_\mathscr{F} \frac{d^2\tau}{\tau_2^2} \, F (\bar \tau )\, \sum_{\tilde m,n} \left( \frac{2\pi i \lambda \bar\eta^3}{\bar\vartheta_1 (\hat\lambda |\bar \tau )}\right)^2\, e^{- \pi \hat\lambda^2 / \tau_2} \, \varLambda_{\tilde m , n} (T,U) \,, \label{topoLag}
\end{equation}
where now $\hat \lambda = \lambda (\tilde m_1 + U \tilde m_2 + \bar\tau (n_1 + U n_2 ))/U_2$, and 
\begin{equation}
\varLambda_{\tilde m , n} (T,U) = T_2\, e^{2 \pi i T (\tilde m_1 n_2 - \tilde m_2 n_1)}\, 
\exp \left\{ - \frac{\pi T_2}{\tau_2 U_2} \left| 
\begin{pmatrix} 1 & \tau \end{pmatrix} 
\begin{pmatrix} \tilde m_1 & \tilde m_2 \\ n_1 & n_2 \end{pmatrix} 
\begin{pmatrix} 1 \\ U \end{pmatrix} \right|^2\right\} \label{22lag}
\end{equation}
is the standard $(2,2)$ lattice in the Lagrangian representation.

Notice  that the amplitude \eqref{topoHam} corresponds to a rank-four heterotic model, also know as the $STU$ model, and thus does not matches verbatim the generating function of \cite{Antoniadis:1995zn}. In fact, the latter corresponds to the rank-three $ST$ model for which the quantum corrections to its type II dual were known up to two loops, and thus a more complete test of heterotic/type II duality could be performed. Clearly, the content of this section and of the following ones can be properly adapted to reproduce the results of  \cite{Antoniadis:1995zn} but, for simplicity, we shall focus on the $STU$ model, based on the standard $T^2 \times K3$ compactification of the heterotic (or type I) string.

\section{The Geometry of the String Background} \label{goem}

The fact that the generating function $G (\lambda ; \tau , \bar\tau )$ may be recast as a functional integral with action
\begin{equation}
S_\text{eff} = \frac{1}{\pi} \int d^2 x\,  \left[ \sum_{i=1,2} \left(\partial Z^i \bar \partial \bar Z^i + \partial \bar Z^i \bar\partial Z^i \right) +\frac{\lambda}{\tau_2} \left( Z^1 \bar\partial Z^2 + \bar Z^2 \bar \partial \bar Z^1 \right)  \right] 
\end{equation}
is suggestive of the fact that the generating function $F (\lambda )$ of the topological amplitudes may be interpreted as the partition function of the heterotic string on some non-trivial background.  Moreover, amplitudes involving an increasing number of external states back-react on the geometry and thus have the effect of modifying the original space-time.

Notice, however, that the $\lambda$-dependent deformation seems to affect only the right-moving coordinates, while one is left with just a functional integral over the bosonic coordinates. Still the heterotic string contains left-moving fermions.\footnote{aside from the right moving ones describing the gauge degrees of freedom whose contribution is encoded in the anti-holomorphic function $F (\bar \tau )$.} Therefore the effective description in terms of the functional integral of   $S_\text{eff}$ emerges after a proper integration over the fermionic coordinates which compensate for the $\lambda$ deformation of the left-moving bosons, as a result of the left-over space-time supersymmetry. As a result, the geometric background is expected to treat symmetrically the bosonic coordinates, while coupling only to the {\em space-time} fermions of the heterotic string. In the following, we shall describe this geometric background, while we will defer the coupling to world-sheet fermions to sections \ref{OSMelvin} and \ref{HSMelvin}. 

The propagation of the bosonic string on a generic background is dictated by the world-sheet action
\begin{equation}
S = \frac{1}{2\pi \alpha '} \int d^2 x \, \left[ \left( G_{MN} (X) + B_{MN} (X) \right) \partial X^M \bar \partial X^N  + \tfrac{1}{4} \alpha ' \sqrt{h} R^{(2)}\, \varPhi (X) \right]\,.
\end{equation}
Consistency then requires that the beta functions associated to the metric field $G_{MN}$, the antisymmetric field $B_{MN}$ and the dilaton $\varPhi (X)$ vanish at any order in $\alpha '$. This clearly puts strong constraints on the background fields and only few geometries are known to yield a consistent two-dimensional CFT. 

In the mid-nineties a thorough study of non-trivial geometries affording an exact CFT description has been undertaken. A particularly interesting (and simple) exact solution is given by the  $\sigma$-model  \cite{Russo:1994cv,Russo:1995tj}
\begin{equation}
\mathscr{L} = \partial \rho \bar\partial \rho + \rho^2 ( \partial \varphi + q \partial y)( \bar\partial \varphi + q \bar\partial y ) + \partial y \bar\partial y + \partial \boldsymbol{x} \bar\partial \boldsymbol{x}\,. \label{1Melvinsigma}
\end{equation}
Here $y$ is a compact coordinate on a circle of radius $R$, $(\rho , \varphi )$ are polar coordinates on a plane and $\boldsymbol{x} = \{x^\alpha \}$ describe the rest of space-time.  

To identify the (lower-dimensional) geometry one uses the KK ansatz 
\begin{equation}
g_{\mu\nu} = G_{\mu \nu} - G_{55} A_\mu A_\nu\,, \qquad A_\mu = G_{\mu 5} /  G_{55} \,, \qquad G_{55} = e^{2\sigma} \,,
\end{equation}
which, combined with eq. \eqref{1Melvinsigma} yields
\begin{equation}
\begin{split}
\text{d}s^2 &= \text{d}\rho^2 + \rho^2 \, F (\rho) \, \text{d} \varphi^2 + \text{d}  \boldsymbol{x}^2 \,,
\\
A &= q \rho^2 F (\rho ) \, \text{d} \varphi\,,
\\
e^{-2\sigma} &= F (\rho ) \,,
\end{split}\label{Melgeom1}
\end{equation}
describing a magnetic flux-tube in a properly curved background, where $F^{-1} (\rho ) = 1 + q^2\, \rho^2$. 

Clearly, one can further generalise this model to include two magnetic flux tubes along different directions. Introducing a second set of polar coordinates, the $\sigma$-model action becomes
\begin{equation}
\begin{split}
\mathscr{L}_1 &= \partial \boldsymbol{x} \bar\partial \boldsymbol{x} + \partial \rho_1 \bar\partial \rho_1 + \rho_1^2 (\partial \varphi_1 + q_1 \partial y ) (\bar\partial \varphi_1 + q_1 \bar\partial y )
\\
& \quad + \partial \rho_2 \bar\partial \rho_2 + \rho_2^2 (\partial \varphi_2 + q_2 \partial y ) (\bar\partial \varphi_2 + q_2 \bar\partial y )
+\partial y \bar \partial y \,. \label{gen2Melvin}
\end{split}
\end{equation}
and the associated four-dimensional geometry now reads
\begin{equation}
\begin{split}
\text{d} s^2 &= \text{d}\rho_1^2 + \rho_1^2 G (1+ q_2^2 \rho_2^2 ) \text{d} \varphi_1^2 + \text{d}\rho_2^2  + \rho_2^2 G (1+ q_1^2 \rho_1^2 ) \text{d} \varphi_2^2 
\\
&\quad - 2 G q_1 q_2 \rho_1^2 \rho_2^2 \text{d}\varphi_1 \text{d}\varphi_2 \,,
\\
A &= G (q_1 \rho_1^2 \text{d}\varphi_1 + q_2 \rho_2^2 \text{d} \varphi_2 )\,,
\\
\phi &=\phi_0\,,
\\
e^{-2\sigma} &= G \,,
\end{split}
\end{equation}
where $G^{-1} = 1 + q_1^2 \rho_1^2 + q_2^2 \rho_2^2$. For weak magnetic flux tubes, the four-dimensional geometry is flat $g_{\mu\nu} = \delta_{\mu\nu} + O (q_i^2)$, the radius of the internal circle constant, while the gauge field describes a constant magnetic field
\begin{equation}
F_{\mu \nu} = 2 \begin{pmatrix} 
0 & -q_1 & 0 & 0  \\
q_1 & 0 & 0 & 0 \\
0 & 0 & 0 & -q_2 \\
0 & 0 & q_2 & 0 \end{pmatrix}\,. 
\label{MelvinFS}
\end{equation}
The choice $q_1 =\pm q_2$ is rather interesting since the constant magnetic field is (anti-)self-dual, $F = \pm \star F$, and thus the background is compatible with the presence of supersymmetry. 

This choice is particularly important for our case, since the graviphotons scattered in the topological amplitudes, corresponding to the off-diagonal components of a higher-dimensional metric, are chosen to be anti-self-dual, as required by the nature of the higher-derivative $F$-terms.

To make full contact with the graviphoton vertex operators involved in the topological amplitudes, which have a leg on the internal $T^2$, the Melvin background can be further generalised to involve a compact $T^2$ with complex structure $U$ and K\"ahler form $T$
\begin{equation}
\begin{split}
\mathscr{L}_2 &= \partial \boldsymbol{x} \bar\partial \boldsymbol{x} + \partial \rho_1 \bar\partial \rho_1 + \rho_1^2 (\partial \varphi_1 + q_i \partial y^i ) (\bar\partial \varphi_1 + q_i \bar\partial y^i )
\\
& \quad + \partial \rho_2 \bar\partial \rho_2 + \rho_2^2 (\partial \varphi_2 - q_i \partial y^i ) (\bar\partial \varphi_2 - q_i \bar\partial y^i )
+\partial y^i \bar \partial y^i \,,
\end{split} \label{ASDMelvin}
\end{equation}
where we have fixed the magnetic flux tubes to be anti-selfdual. 

\section{Strings on Melvin Space} \label{OSMelvin}

Let us go back to the simple $\sigma$-model described by the Lagrangian \eqref{1Melvinsigma}. As already anticipated it corresponds to an  exact CFT, and a very simple one. Indeed, the field redefinition $\varphi \to \varphi_0 = \varphi + q y$ drastically simplifies the Lagrangian
\begin{equation}
\mathscr{L} = \partial \rho \bar\partial \rho + \rho^2 \partial \varphi_0 \bar\partial \varphi_0 + \partial y \bar\partial y\,, \label{Melvinflat}
\end{equation}
which actually describes free fields. Clearly, this model is not equivalent to a flat Euclidean geometry because of the compactness of the $y$ coordinate and the fact that $\varphi_0$ is no longer an angular variable, $\varphi_0 \sim \varphi_0 + 2 \pi qR$. As a result, the complex coordinate $Z_0 = \rho e^{i\varphi_0}  = e^{i q y}\, Z$ picks-up a non-trivial phase whenever the string winds around the compact circle
\begin{equation}
Z_0 (\tau , \sigma + \pi ) = e^{2 i \pi n q R} \, Z_0 (\tau , \sigma ) \qquad \text{as} \qquad y (\tau , \sigma + \pi ) = y (\tau , \sigma ) + 2 \pi R n \,.
\label{Melvinorb}
\end{equation}
 The flat sigma model \eqref{Melvinflat} together with the identification \eqref{Melvinorb} has a close resemblance with (freely acting) orbifold compactifications, which is behind the simplicity and the exact CFT descriptions of the  model. 
 
The quantisation of strings on the Melvin background proceeds as usual, and we shall not indulge here on a long and detailed description \cite{Russo:1994cv,Russo:1995tj}. It suffices to say that, because of the mixing of the true angular variable $\varphi$ and the compact coordinate $y$ the quantised conjugate momenta are
\begin{equation}
\begin{split}
P_\varphi &= \frac{1}{ 2\pi \alpha '} \int_0^\pi d\sigma \, \rho^2 \left( \dot \varphi + q \dot y \right) = L \,, 
\\
P_y &= \frac{1}{2\pi \alpha '} \int_0^\pi d\sigma\, \dot y + \frac{q}{ 2\pi \alpha '} \int_0^\pi d\sigma \, \rho^2 \left( \dot \varphi + q \dot y \right) = \frac{k}{R}\,,
\end{split} \label{Melmomenta}
\end{equation}
with $L, k\in\mathbb{Z}$ being the angular and Kaluza-Klein momenta. As a result, the zero-mode part of $y$, which has standard integer frequencies, is
\begin{equation}
y (\tau , \sigma ) = y_0 + 2 \alpha ' \left( \frac{k}{R} - q L \right) \, \tau + 2 n R \, \sigma + \text{oscillators} \,.
\end{equation}

As for the $Z$ coordinate, states in the zero-winding sector have the conventional integer-mode expansion while, whenever $n\not =0$, frequencies are shifted, as pertains to a twisted coordinate,
\begin{equation}
\begin{split}
Z_0 (\sigma , \tau ) &= \sqrt{\alpha '} \left[ \sum_{m=1}^\infty \frac{a_m}{\sqrt{m-\nu}} e^{-2 i (m-\nu )\sigma_+} + \sum_{m=0}^\infty \frac{b^\dagger_m}{\sqrt{m+\nu}}e^{-2i (m+\nu ) \sigma_+}\right.
\\
& \qquad  \left.+ \sum_{m=1}^\infty \frac{\tilde a _m}{\sqrt{m+\nu}} e^{-2i (m+\nu)\sigma_-} + \sum_{m=0}^\infty \frac{\tilde b^\dagger_m}{\sqrt{m-\nu}}e^{-2i(m-\nu)\sigma_-}\right] \,.
\end{split}
\end{equation}
Here  $\nu = qR n - [qR n]$, for $n$ positive, and  $\nu = q R  n - [q R n]+1$ for $n$ negative. 

The remaining spectator coordinates have the standard mode expansion, so that the full Hamiltonian reads
\begin{equation}
L_0+\tilde L_0 = N+\tilde N - \nu (J-\tilde J) +\frac{\alpha' }{2}\left[ \left( \frac{nR}{\alpha'}\right)^2 + \left( \frac{k}{R} - q ( J + \tilde J )\right)^2 \right] - \frac{1}{4} (1-2\nu)^2 \,.
\end{equation}
Here $N$ and $\tilde N$ are the total left-moving and right-moving number operators, while 
\begin{equation}
J+\tilde J = \tilde a^\dagger_0 \tilde a_0 - b^\dagger_0 b_0 + \sum_{m=1}^\infty \left( \tilde a^\dagger_m \tilde a_m + a^\dagger_m a_m - \tilde b^\dagger_m \tilde b_m - b^\dagger_m b_m \right)
\end{equation}
refers to the angular momentum on the Melvin plane, conjugate to the angular variable $\varphi$. Notice that, as anticipated, for zero winding number, the coordinate $Z$ involves a non-trivial zero mode which contributes to the angular momentum via the standard orbital part. In this case, the oscillators $b_0$ and $\tilde a_0$, together with their hermitian conjugates, are absent. This is expected, since the latter generate the tower of Landau levels associated to the closed-string graviphoton. Their contribution to the partition function
\begin{equation}
\mathscr{Z} = \int_\mathscr{F} \frac{d^2\tau}{\tau_2}\, \frac{R}{\pi \alpha ' \tau_2} \sum_{\tilde k , n \in \mathbb{Z}} e^{-\pi \frac{R^2}{\alpha ' \tau_2} |\tilde k + \tau n|^2}\, \text{Tr}_\nu \left[ e^{2 i \pi R q \tilde k (J+\tilde J) }\, q^{L_0} \, \bar q^{\bar L_0}\right]\,,
\label{part1Melvin}
\end{equation}
is 
\begin{equation}
\begin{split}
\int d^2 p \, \langle p | e^{2 i \pi L R B_\text{c} \tilde k} |p\rangle\, (q\bar q)^{\frac{\alpha '}{4} p^2 }  &= \int d^2 p \, \langle p | \theta \cdot p \rangle 
\\
&= \frac{1}{\det (1-\theta )} 
\\
&= \frac{1}{ \left(2\, \sin (\pi L RB_\text{c} \tilde k )\right)^2} \,,
\end{split}
\end{equation}
when $\tilde k \not =0$, and $\tau_2^{-1}$ for $\tilde k =0$.

Upon evaluating the trace in \eqref{part1Melvin} one gets the final result
\begin{equation}
\mathscr{Z} \propto 2\pi R \int_\mathscr{F} \frac{d^2 \tau}{\tau_2^{13}} \, \sum_{\tilde k , n\in \mathbb{Z}} e^{-\pi \frac{R^2}{\alpha ' \tau_2} |\tilde k + \tau n|^2}\, \mathscr{Z} (\tilde k , n )\,, \label{Melbosons1}
\end{equation}
with
\begin{equation}
\begin{split}
\mathscr{Z} (0,0) &= \frac{1}{\tau_2} \, \frac{1}{(\eta \bar \eta )^{24}} \,,
\\
\mathscr{Z} (\tilde k , n ) &= \frac{1}{(\eta \bar \eta )^{24}} \, \left| \frac{ \vartheta_1^\prime (0|\tau )}{2\pi\, \vartheta_1 (\chi |\tau )}\right|^2 \, e^{-\frac{\pi}{2\tau_2} (\chi-\bar\chi )^2}\,, 
\end{split}\label{Melbosons2}
\end{equation}
where $\chi = q R (\tilde k + n \tau)$, and we have omitted in $\mathscr{Z}$ the standard multiplicative prefactor involving the (infinite) volume of the non-compact space and powers of $\alpha '$. 

Alternatively, the partition function can be obtained by evaluating the Gaussian functional integral and, in this case, the non-holomorphic factor $e^{-\frac{\pi}{\tau_2} (\chi - \bar\chi )^2}$ follows from a $\zeta$-function regularisation of the functional determinant. 

The connection of the Melvin deformation with freely acting orbifolds is rather transparent in the case where the Melvin twist is a rational number. In this case, using the periodicity properties \eqref{JThPeriodicity} of the Jacobi theta functions, one can directly Poisson re-sum over the windings $\tilde k$ and write the torus amplitude in the Hamiltonian representation. The result is then a {\em conventional} orbifold where the rational twists are accompanied by fractional shifts along the compact circle. As an illustrative example, let us consider the simple case $qR=\frac{1}{2}$. Using
\begin{equation}
\vartheta_1 (1/2|\tau) \sim \vartheta_2 (0|\tau) \,, \ \vartheta_1 (\tau /2|\tau ) \sim \vartheta_4 (0|\tau )\,, \ \vartheta_1 ((1+\tau)/2|\tau) \sim \vartheta_3 (0|\tau )\,,
\end{equation}
one gets the following expression for the partition function:
\begin{equation}
\begin{split}
\mathscr{Z} &= \int_\mathscr{F} \frac{d^2\tau}{\tau_2^{25/2}} \, \frac{1}{(\eta\bar\eta )^{22} }\, \sum_{m,n\in\mathbb{Z}} \left[ \frac{1}{\tau_2} \frac{1}{|\eta |^4} \varLambda_{m,n} (2R ) + \left|\frac{2\eta}{\vartheta_2}\right|^2 \, (-1)^m\, \varLambda_{m,n} (2R) \right.
\\
&\qquad \qquad \left.+ 4\, \left|\frac{\eta}{\vartheta_4}\right|^2 \,\varLambda_{m,n+\frac{1}{2}} (2R) + 4\left|\frac{\eta}{\vartheta_3}\right|^2 \, (-1)^m\, \varLambda_{m,n+\frac{1}{2}} (2R) \right] \,,
\end{split}
\end{equation}
where $\varLambda_{m,n+\alpha} (2R) = q^{\frac{1}{4} \left(\frac{m}{2R} + (n+\alpha) 2R\right)^2}\, \bar q^{\frac{1}{4} \left(\frac{m}{2R} - (n+\alpha) 2R\right)^2}$. This partition function indeed corresponds to the freely acting $(\mathbb{C}\times S^1 (2R)) /\mathbb{Z}_2$ orbifold, where the inversion $z\to -z$ of the $\mathbb{C}$ coordinate is accompanied by a shift along $S^1(2R) $ by half of its circumference.

The construction can be easily generalised to include diverse flux tubes along various directions as in \eqref{gen2Melvin}: it suffices to add in  $\mathscr{Z} (\tilde k , n ) $ of eq.~\eqref{Melbosons2} extra $\chi_i$-dependent terms, with $\chi_i = q_i R ( \tilde k + n \tau) $ associated to the $i$-th flux-tube, while properly taking into account the zero-mode contribution, {\em i.e.} the powers of $\tau_2$, in $\mathscr{Z}$ and in $\mathscr{Z} (0,0)$.

Adding fermions to the previous discussion is also straightforward \cite{Russo:1995ik}. In the RNS formulation the world-sheet action generalises to
\begin{equation}
\begin{split}
\mathscr{L}_\text{RNS} &= \tfrac{1}{2}\left(  (\partial + i q \partial y) Z (\bar\partial - i q \bar\partial y)  Z^* + \text{c.c.}\right) \, + \lambda_\text{R}^* (\partial +i q \partial y ) \lambda_\text{R} +  \lambda_\text{L}^* (\bar\partial -i q \bar\partial y ) \lambda_\text{L} 
\\
&\qquad + \partial y \bar\partial y  + \psi_\text{R} \partial \psi_\text{R} + \psi_\text{L} \bar\partial \psi_\text{L} \,,
\end{split}
\end{equation}
where the complex fermions $\lambda_\text{L,R}$ are the partners of the complex coordinate $Z$, while $\psi_\text{L,R}$ are real and are in correspondence with $y$. Also in this case, the change of variables $Z_0 = e^{i q y} Z$ and $\lambda_{0,\text{R}(\text{L})} = e^{\pm i q y}\, \lambda_{0,\text{R}(\text{L})}$ casts the RNS action in the simple form
\begin{equation}
\mathscr{L}_\text{RNS} = \tfrac{1}{2} \, \left( \partial Z_0 \, \bar\partial Z^*_0 + \partial Z_0^* \, \bar\partial Z_0 \right) + \lambda_{0,\text{R}}^* \partial \lambda_{0,\text{R}} +  \lambda_{0,\text{L}}^* \, \bar\partial \lambda_{0,\text{L}} + \partial y \, \bar\partial y +  \psi_\text{R} \partial \psi_\text{R} + \psi_\text{L} \bar\partial \psi_\text{L}  \,,
\end{equation}
which only involves free fields, but with twisted boundary conditions
\begin{equation}
Z_0 (\tau , \sigma +\pi ) = e^{2 i \pi q R n} Z_0 (\tau , \sigma)\,, \quad
\lambda_{0,\text{R}(\text{L})} (\tau , \sigma +\pi )= \pm e^{\pm 2 i \pi q R n} \lambda_{0,\text{R}(\text{L})}  (\tau , \sigma ) \,.
\label{twistedMfields}
\end{equation}
The overall $\pm$ sign in the periodicity condition for the world-sheet fermions refers, as usual, to the R ($+$) and NS ($-$) sectors. 

The quantisation of the action $\mathscr{L}_\text{RNS}$ follows the standard procedure and the contribution to the partition function of a complex non-chiral twisted fermion with right-moving spin structure $(a , b )$ and left-moving spin structure $(c , d )$ is
\begin{equation}
\frac{\theta \big[ {\textstyle{a \atop b}}\big]  (\chi |\tau ) }{\eta}\, \frac{\bar \theta \big[ {\textstyle{c \atop d}}\big]  (\bar\chi | \bar\tau ) }{\bar\eta}\, e^{\frac{\pi}{2\tau_2} (\chi-\bar\chi )^2}\,. \label{Melfermi}
\end{equation}
Notice that the determinant of the Dirac operator does not yield holomorphic factorisation, the obstruction being related to the Quillen anomaly \cite{Quillen:1985, AlvarezGaume:1986uh, Belavin:1986cy} associated to the two-dimensional flat connection $\chi$.  

The full partition function is then a sum over spin structures and depends on the choice of GSO projection. For the type IIA/IIB superstrings reads
\begin{equation}
\mathscr{Z} \propto 2\pi R \int_\mathscr{F} \frac{d^2 \tau}{\tau_2^{5}} \, \sum_{\tilde k , n\in \mathbb{Z}} e^{-\pi \frac{R^2}{\alpha ' \tau_2} |\tilde k + \tau n|^2}\, \mathscr{Z} (\tilde k , n )\,,
\end{equation}
with
\begin{equation}
\begin{split}
\mathscr{Z} (0,0) &= \frac{1}{\tau_2} \, \frac{1}{(\eta \bar \eta )^{12}} \, \left|\tfrac{1}{2} \sum_{a , b} \eta_{a,b}^\text{A(B)}\theta ^4 \big[ {\textstyle{a \atop b}}\big]  (0 |\tau ) \right|^2
 \,,
\\
\mathscr{Z} (\tilde k , n ) &= \frac{1}{(\eta \bar \eta )^{12}} \, \left| \frac{ \vartheta_1^\prime (0|\tau )}{2\pi\, \vartheta_1 (\chi |\tau )}\right|^2 \, 
\left|\tfrac{1}{2} \sum_{a , b} \eta_{a,b}^\text{A(B)}\theta\big[ {\textstyle{a \atop b}}\big]  (\chi |\tau ) \theta^3\big[ {\textstyle{a \atop b}}\big]  (0 |\tau ) \right|^2 \,,
\end{split}
\end{equation}
where $\eta_{a,b}^\text{A(B)}$ are the standard GSO phases for type IIA or IIB superstrings.

For the heterotic string, things are slightly more complicated since one has to decide whether or not the internal (gauge) degrees of freedom couple or not to the graviphotons.\footnote{See \cite{Russo:1995aj} for a discussion of different cases and the analysis of the associated massless excitations.} We shall be more explicit on this point in the next section.

Notice that the Melvin deformation acting on a single two-plane does not preserve any space-time supersymmetry and, in addition, induces instabilities due to the emergence of tachyonic excitations for suitable choices of the magnetic background. This situation can be improved if magnetic fields are turned on along different directions \cite{Russo:2001na}. This should not be a surprise after all, since the graviphoton backgrounds have a more natural description in terms of {\em orbifold} twists of the complex coordinates \eqref{Melvinorb} and we know that, in this case, some supersymmetries are preserved if \cite{Dixon:1985jw,Dixon:1986jc}
\begin{equation}
\sum_i q_i =0\,. \label{MelvSUSYcond}
\end{equation}

\section{Heterotic String on Melvin Space}\label{HSMelvin}

In its fermionic formulation, the heterotic string involves ten non-chiral bosons $x^M$ describing the embedding of the world-sheet in the ten-dimensional space-time, ten real right-moving fermions $\lambda^M_\text{R}$ superpartners of the chiral $x_\text{R}^M$, thus yielding $\mathscr{N}=(0,1)$ supersymmetry in $d=2$, together with 32 real left-moving fermions $\zeta_\text{L}^I$, describing the gauge degrees of freedom. 

The geometric background \eqref{Melgeom1} describes an exact CFT also for the heterotic string. The quantisation of the world-sheet bosons is straightforward and its contribution to the partition function is encoded in eqs. \eqref{Melbosons1} and \eqref{Melbosons2} with a suitable change in the dimensionality of the target space.

Moving to the fermions there are now more options and some subtleties. The simplest model is when the Melvin background couples to a pair of $\lambda_\text{R}$'s and to a pair of $\zeta_\text{L}$ fermions from the gauge degrees of freedom. This is reminiscent of the standard embedding in orbifold compactifications, and the relevant contribution to the two-dimensional Lagrangian is
\begin{equation}
\mathscr{L} \supset \lambda_\text{R}^* (\partial +i q \partial y ) \lambda_\text{R} +  \zeta_\text{L}^* (\bar\partial -i q \bar\partial y ) \zeta_\text{L} \,.
\end{equation}
The integration over $\lambda_\text{R}$ and $\zeta_\text{L}$ is again Gaussian and evaluates to the determinant of the {\em non-chiral} Dirac operator in the presence of the flat Melvin connection. The result is a modular-invariant combination of the contributions \eqref{Melfermi} with proper GSO projections in the (right-moving) supersymmetric and (left-moving) gauge sectors. 

The second option where the Melvin background only deforms the contribution of the space-time fermions, and not that of the gauge degrees of freedom, is more subtle since it requires the evaluation of the determinant of a chiral Dirac-Weyl operator with a flat connection, {\em i.e.} with a Melvin deformation. From the point of view of the two-dimensional world-sheet the chiral coupling with the flat connection induces a gauge anomaly which translates into a non-holomorphic expression for the chiral determinat \cite{AlvarezGaume:1986uh,Griguolo:356541}
\begin{equation}
\det (\partial + \chi) = e^{-\frac{\pi}{\tau_2} |\chi|^2}\, e^{\frac{\pi}{2\tau_2}\chi^2} \frac{\theta \big[{\textstyle{a\atop b}}\big] (\chi |\tau)}{\eta (\tau )}\,,
\label{chiraldet}
\end{equation}
for a pair of spin-structure $(a,b)$.

Putting together the contributions from all degrees of freedom, one finds the partition function
\begin{equation}
\mathscr{Z}_\text{Het} = 2\pi R \int_\mathscr{F} \frac{d^2\tau}{\tau_2^5} \bar{\mathscr{Z}}_\text{gauge} (\bar \tau )\, \sum_{\tilde k , n\in\mathbb{Z}} e^{-\pi \frac{R^2}{\alpha ' \tau_2} |\tilde k + \tau n|^2} \mathscr{Z}_\text{B} (\tilde k , n )\, \mathscr{Z}_\text{F} (\tilde k , n)\,.
\end{equation}
for the heterotic string on the Melvin background. 
Here $\bar{\mathscr{Z}}_\text{gauge} (\bar \tau )$ encodes the contribution of the gauge degrees of freedom for the $\text{SO} (32)$ or $\text{E}_8 \times \text{E}_8$ groups,  $\mathscr{Z}_\text{B} (\tilde k , n )$ encodes the contribution of the bosonic coordinates $x^M$ and is given by eq. \eqref{Melbosons2} with the suitable replacement of $24 \to 8$ in the number of $\eta\, \bar \eta$ functions in the denominator, while 
\begin{equation}
\begin{split}
\mathscr{Z}_\text{F} (\tilde k ,n) &=e^{-\frac{\pi}{\tau_2} |\chi |^2}\, \frac{e^{ \frac{\pi}{2\tau_2} \chi^2}}{2 \eta^4} \sum_{a ,b} \eta_{a , b}\, \theta \big[ {\textstyle{a \atop b}}\big] (\chi | \tau ) \, \theta^3 \big[ {\textstyle{a \atop b}}\big] (0 | \tau )   
\\
&=  e^{\frac{\pi}{2\tau_2} \chi (\chi - 2 \bar\chi )}\, \frac{\vartheta_1^4 (\tfrac{1}{2}\chi |\tau )}{\eta^4} \,,
\end{split}
\end{equation}
encodes the contribution of the ten world-sheet fermions $\lambda^M_\text{R}$ from the supersymmetric sector, out of which two couple minimally to the flat connection $\chi$. In the last expression we have made use of the Jacobi identity \eqref{RiemannID}.

Using the modular transformations of the $\eta$ and $\theta$ functions  under the $T$ and $S$  transformations in appendix \ref{appJacobi}, it is straightforward to show that the partition function is modular invariant. However, because of the holomorphic anomaly of the chiral determinant \eqref{chiraldet}, ${\mathscr Z}$ is not invariant under the shifts $\chi \to \chi + m + n \tau$.

\section{Heterotic string on $\text{Melvin}\times T^2 \times K3$}

We have now all the ingredients to discuss the heterotic string background which reproduces the generating function of the topological amplitude. The reduced number of space-time supersymmetries implies that the internal space must have reduced holonomy. The unique solution is $T^2 \times K3$ which indeed preserves $\mathscr{N}=2$ supersymmetries in four-dimensions, and we shall work at the orbifold point $K3 = T^4 / \mathbb{Z}_N$ with standard embedding in the gauge sector, although more general configurations can be envisaged. The non-compact flat Euclidean space is then replaced by the Melvin background where all four coordinates are affected by the flux tube. In order to preserve supersymmetry the Melvin deformations, $q_1$ and $q_2$, on the two two-planes of $\mathbb{R}^4$ must obey the condition \eqref{MelvSUSYcond}. Indeed this choice, together with the $K3=T^4/\mathbb{Z}_N$ twist, preserves four of the sixteen components of the ten-dimensional supercharge, since 
\begin{equation}
(s_1 , s_2 , s_3 , s_4 , s_5 ) \to  e^{2 i \pi \left( s_1 q_1 + s_2 q_2 + s_4 \frac{k}{N} - s_5 \frac{k}{N} \right)}\, (s_1 , s_2 , s_3 , s_4 , s_5 )\,,
\end{equation}
where $s_i = \pm 1/2$, and the $(1,0)$ supercharge has, say, an even number of negative helicity states, and $q_1 = - q_2$. This leads to four invariant states corresponding to $s_1=s_2=\pm 1/2$, $s_4=s_5=\pm 1/2$ and $s_3=1/2$.

Notice that the condition $q_1 = - q_2$ on the Melvin twists translates into an anti-self-duality condition (at the linearised level) on the magnetic background \eqref{MelvinFS}, which reflects the anti-self-duality of the graviphoton vertex entering the topological amplitude.

The associated partition function can then be computed combining standard results from orbifold compactifications with those discussed in the previous sections. The bosonic coordinates along the orbifolded $K3$ contribute with 
\begin{equation}
16\, \sin^4 (\pi g /N)\,\left| {\frac{\eta^2}{\vartheta_1 (\xi_N |\tau ) \, \vartheta_1 (-\xi_N |\tau )}}\right|^2 \, e^{-\frac{\pi}{\tau_2} (\xi_N - \bar\xi_N)^2} \,, \qquad \xi_N = \frac{g + h \tau}{N}\,, 
\end{equation}
while the zero modes of the coordinates along the $T^2$ with complex structure $U$ and K\"ahler form $T$ yield the standard winding contribution \eqref{22lag}. To conclude with the bosonic degrees of freedom, those corresponding to the non-compact space-time coordinates contribute with
\begin{equation}
\left| {\frac{\eta^2}{\vartheta_1 (\chi |\tau ) \, \vartheta_1 (-\chi |\tau )}}\right|^2 \, e^{-\frac{\pi}{\tau_2} (\chi - \bar\chi)^2} \,, \label{HetMelvSTBos} 
\end{equation}
where now  $\chi = q (\tilde m_1 + U \tilde m_2 + \tau (n_1 + U n_2 ))/U_2$, and we have coupled the Melvin deformation $q$ to the complex coordinate of the $T^2$. 

Turning to the contribution of the world-sheet fermions, the $\lambda_\text{R}$'s split into two groups, depending whether they point to the Melvin or $K3$ directions. Similarly, the 32 spinors $\zeta_\text{L}$  split into two groups, depending whether they feel or not the standard embedding of the $K3$. Denoting by $(a , b)$ the spin structure of the right-moving spinors and by $(k,l)$ the spin structure of the left-moving ones, one finds
\begin{equation}
\begin{split}
&\frac{\theta \big[{\textstyle {a \atop b}}\big] (\chi |\tau )\, \theta \big[{\textstyle {a \atop b}}\big] (- \chi |\tau )
\theta \big[{\textstyle {a \atop b}}\big] (\xi_N |\tau )\, \theta \big[{\textstyle {a \atop b}}\big] (- \xi_N |\tau )}{\eta^4}
\\
&\qquad \times \frac{\bar\theta \big[{\textstyle {k \atop l}}\big] (\bar\xi_N |\bar\tau )\, \bar\theta \big[{\textstyle {k \atop l}}\big] (- \bar\xi_N |\bar\tau ) \bar\theta^{14} \big[{\textstyle {k \atop l}}\big] (0|\bar\tau)}{\bar\eta^{16}}\, e^{\frac{\pi}{\tau_2} (\xi_N - \bar\xi_N)^2}\, e^{\frac{\pi}{\tau_2}\, \chi (\chi - 2 \bar\chi)} \,.
\end{split}
\end{equation}
We can now combine the various contributions and, upon summation over the orbifold sectors $(g,h)$, over the space-time spin structures with the GSO phase $\eta_{\alpha , \beta}$ and over the spin-structures of the internal gauge degrees of freedom for the $\text{SO} (32)$ gauge group, and after using the Jacobi identity \eqref{RiemannID}, the full partition function reads
\begin{equation}
\mathscr{Z} = - 2\pi T_2 \, \int_\mathscr{F} \frac{d^2\tau}{\tau_2^2} \bar F (\bar\tau )\, \sum_{\tilde m , n} \varLambda_{\tilde m, n} (T,U) \, \left( \frac{\bar\vartheta^\prime_1 (0|\bar\tau )}{2\pi \bar\vartheta_1 (\bar\chi |\bar\tau )}\right)^2\, e^{-\frac{\pi}{\tau_2} \bar\chi^2}\,, \label{pfhetMelK3}
\end{equation}
where
\begin{equation}
\bar F (\bar\tau ) = \frac{1}{2N} \sum_{g,h=0}^{N-1} 16\, \sin^4 (\pi g /N) \, \sum_{k,l=0,\frac{1}{2}} \frac{\bar\theta \big[{\textstyle{k\atop l}}\big] (\bar\xi_N |\bar\tau )\, \bar\theta \big[{\textstyle{k\atop l}}\big] (-\bar\xi_N |\bar\tau ) \bar\theta^{14} \big[{\textstyle{k\atop l}}\big] (0 |\bar\tau )}{\bar\vartheta_1^2 (\bar\xi_N |\bar\tau )\, \bar\eta^{18}}\,.
\end{equation}
This expression clearly reproduces the generating function of the topological amplitudes \eqref{topoLag}, aside from an overall normalisation constant, and identifies the Melvin parameter $q$ with the graviphoton polarisation $\lambda$. One can then Poisson re-sum over the windings $\tilde m$ to get the alternative {\em Hamiltonian} representation which reproduces eq. \eqref{topoHam}. The origin of the holomorphic anomaly of the topological amplitudes is now traced back to the lack of the quantum gauge invariance on the world-sheet. 

The sum over the spin structures $a$ and $b$ does not yield a vanishing partition function \eqref{pfhetMelK3} as, instead, one might have naively expected since the background preserves four supercharges. In fact, the non vanishing of \eqref{pfhetMelK3} is not at all in contradiction with a residual supersymmetry, and it is quite simple to understand why. Barring the coupling of the Melvin twist with the compact coordinate of the (otherwise) spectator $T^2$, one may think of this background as a two-dimensional compactification on $K3 \times K3$ which would preserve $\mathscr{N} = (4,0)$ supersymmetries in two dimensions.\footnote{To be more precise, this compactification would correspond to a background of the type $\mathbb{R}^2 \times T^4/\mathbb{Z}_N \times T^4/\mathbb{Z}_M$. However, for our purposes this is at all equivalent to replacing $ T^4/\mathbb{Z}_N$ with its non-compact version $\mathbb{C}^2 / \varGamma_q$, where $\varGamma_q$ defines a continuous Abelian rotation of  angles $q$ and $-q$ on the two complex planes.} Now representations of the two-dimensional supersymmetry algebra are rather peculiar since in $d=2$ there is a Lorentz invariant notion of whether massless particles move towards left or right. Therefore, chiral linear supersymmetry does not imply an equality of bosonic and fermionic degrees of freedom, and indeed there exist representations with only fermionic (or bosonic) excitations \cite{Hull:1985jv}. This explains the non-vanishing of the partition function \eqref{pfhetMelK3}.\footnote{See \cite{Dasgupta:1996yh, Florakis:2017zep} for other instances of two-dimensional supersymmetric compactifications with non-vanishing partition function.} 

\subsection{The $\mathscr{N}=2^*$ case}

The generalisation to the case of $\mathscr{N}=2^*$ theories is rather straightforward. To start, one should notice that in order to keep in the spectrum a massive hypermultiplet in the adjoint representation of the gauge group, the breaking $\mathscr{N}=4 \to \mathscr{N}=2$ cannot be hard, but rather spontaneous. In this way, by tuning the (partial) supersymmetry breaking scale, one can interpolate correctly between the original $\mathscr{N}=4$ theory and the final $\mathscr{N}=2$ one.  In string theory, spontaneous symmetry breaking is best achieved in terms of freely acting orbifolds, whereby the orbifold twist which reproduces the $K3$ surface is accompanied by a shift along an extra compact direction. In heterotic string, however, modular invariance of the one-loop partition function requires a simultaneous twist of the gauge degrees of freedom, and this leads automatically to different gauge group representations for the vector and the massive hypermultiplet.  In order to avoid this unpleasant feature, in \cite{Florakis:2015ied} only the left-moving internal bosonic coordinates were twisted, along with their fermionic superpartners, while the right-moving sector is left untouched by the freely-acting orbifold. Clearly, this action corresponds to an asymmetric orbifold and thus the model is defined only at the special SO(8) point of the $T^4$ lattice, for an asymmetric $\mathbb{Z}_2$ action. The inclusion of the Melvin deformation proceeds as before. As a result, the non-compact space-time bosonic coordinates contribute to the one-loop partition function with the combination \eqref{HetMelvSTBos}, the left-moving fermions with spin structure $(a,b)$ contribute with the combination
\begin{equation}
\frac{\theta \big[{\textstyle {a \atop b}}\big] (\chi |\tau )\, \theta \big[{\textstyle {a \atop b}}\big] (- \chi |\tau )
\theta \big[{\textstyle {a \atop b}}\big] (\xi_2 |\tau )\, \theta \big[{\textstyle {a \atop b}}\big] (- \xi_2 |\tau )}{\eta^4}\, e^{\frac{\pi}{\tau_2} \, \xi_2 (\xi_2 - 2 \bar\xi_2)}\, e^{\frac{\pi}{\tau_2}\, \chi (\chi - 2 \bar\chi)} \,,
\end{equation}
while the $(4,4)$ asymmetrically twisted lattice contributes with
\begin{equation}
\frac{\eta^6\, e^{-\frac{\pi}{\tau_2} \, \xi_2 (\xi_2 - 2 \bar\xi_2)}}{\vartheta_1 (\xi_2 |\tau )\vartheta_1 (-\xi_2 |\tau )}\,  \left( \sum_{\gamma , \delta} (-1)^{g(\gamma +h)} \bar\theta ^4\big[{\textstyle {\gamma \atop \delta}}\big] - (-1)^{g}\, \bar\theta^4 \big[{\textstyle {1/2+h/2 \atop 1/2+g/2}}\big] \right)\,.
\end{equation}
The zero-modes of the $T^2$ lattice and the gauge degrees of freedom contribute with the combination
\begin{equation}
\Lambda_{\tilde m , n} \big[{\textstyle {h \atop g}}\big]  (T,U ) \, \bar\chi_8^2 (\bar\tau) \,,
\end{equation}
$\bar\chi_8$ being the character associated to the affine level-one $E_8$ group. Notice that, because of the free action of the asymmetric orbifold twist, the $\tilde m_1$ and  $n^1$ winding numbers  are now shifted, $ (\tilde m^1 , n^1)  \to ( \tilde m^1 +g/2 ,  n^1 +h/2)$, also in the Melvin deformation $\chi$. 

As usual, the Jacobi identity \eqref{RiemannID} drastically simplifies the final expression of the free energy, which reads
\begin{equation}
\mathscr{Z} = - 2 \pi T_2 \int_\mathscr{F} \frac{d^2\tau}{\tau_2^2}\, \sum_{h,g=0,1} \bar H \big[{\textstyle {h \atop g}}\big]  (\bar \tau )\, \sum_{\tilde m , n}  \varLambda_{\tilde m , n}  \big[{\textstyle {h \atop g}}\big]  (T,U) \left( \frac{\bar\vartheta_1^\prime (0|\bar\tau )}{2\pi \bar \vartheta_1 (\bar\chi |\bar\tau )}\right)^2 \, e^{-\frac{\pi}{\tau_2} \bar\chi^2}\,.
\end{equation}
with
\begin{equation}
\bar H \big[{\textstyle {h \atop g}}\big]  (\bar \tau )= \sum_{\gamma , \delta} (-1)^{g(\gamma +h)} \bar\theta ^4\big[{\textstyle {\gamma \atop \delta}}\big] - (-1)^{g}\, \bar\theta^4 \big[{\textstyle {1/2+h/2 \atop 1/2+g/2}}\big] \,.
\end{equation}
This partition function reproduces the topological amplitude of \cite{Florakis:2015ied} after Poisson summation over the windings $\tilde m_i$ is performed. Clearly, the Nekrasov free energy for $\mathscr{N}=2^*$ gauge theories is reproduced in the field theory limit.

\section{Open Strings on Melvin Space}

The case of open strings follows similar steps \cite{Dudas:2001ux,Takayanagi:2001aj,Angelantonj:2002id}. Assuming NN boundary conditions along the Melvin planes and the compact $S^1$, only states carrying Kaluza-Klein momenta are present. As a result, both the bosonic and fermionic coordinates have unshifted frequencies, thus corresponding to standard untwisted coordinates. The Kaluza-Klein momentum is shifted as in \eqref{Melmomenta} and the total  Hamiltonian  reads
\begin{equation}
H_\text{open} = N + \alpha ' \left( \frac{k}{R}- q J \right)^2 + \varDelta\,,
\end{equation}
where $N$ is the full oscillator number, $J$ is the angular momentum associated to the Melvin plane and $\varDelta$ is the vacuum energy in the NS or R sectors. 

The annulus partition function for a D$p$-brane can be computed after Poisson summation over the Kaluza-Klein momenta, and reads
\begin{equation}
\mathscr{A} =  2\pi R \, \int_0^\infty \frac{dt}{t^{(p+1)/2}} \sum_{\tilde k \in\mathbb{Z}} e^{-\pi (R \tilde k)^2 / \alpha ' t}\, \mathscr{A} (\tilde k )\,, 
\label{AsingleB}
\end{equation}
with
\begin{equation}
\begin{split}
\mathscr{A} (0) &= \frac{V_2}{4 \pi^2 \alpha '}\, \frac{1}{t}\, \sum_{a,b} \tfrac{1}{2} \, \eta_{a,b}\, \frac{\theta^4 \big[ {a \atop b}\big] (0) }{\eta^{12}}\,,
\\
\mathscr{A} (\tilde k\not= 0 ) &=\frac{1}{2\, \sin (\pi qR  \tilde k)} \, \sum_{a,b} \tfrac{1}{2} \,\eta_{a,b}\, \frac{ \theta \big[ {a \atop b}\big] (qR \tilde k  )\,  \theta^3 \big[{a\atop b}\big] (0 )}{\vartheta_1 (qR\tilde k )\, \eta^9 } \,,
\end{split}\label{AnnMelvin}
\end{equation}
where we have omitted the explicit dependence of the theta and eta functions on the modulus $\frac{1}{2} i t$ of the double covering torus.
Notice that the pre-factor in $\mathscr{A} (\tilde k )$ is a combination of the integration over the momentum along the Melvin plane and the correct counting of bosonic oscillators,
\begin{equation}
\frac{1}{ (2\, \sin (\pi qR  \tilde k) )^2} \, \frac{ 2 \sin (\pi qR \tilde k) }{\vartheta_1 (qR\tilde k |\tau)}\,.
\end{equation}

Upon an $S$ modular transformation, the one-loop amplitude for open strings is mapped to 
\begin{equation}
\tilde{\mathscr{A}} = 2^{-(p+1)/2} \, \int_0^\infty d\ell \, \ell^{(p-9)/2}\, \sum_{n\in\mathbb{Z}}\, e^{-\frac{\pi\ell}{2\alpha'} (nR)^2}\, \tilde{\mathscr{A}} (n)\,,
\end{equation}
with
\begin{equation}
\begin{split}
\tilde{\mathscr{A}} (0) &= \frac{1}{2}\, \frac{V_2}{4\pi^2 \alpha '}\,   \sum_{a,b} \tfrac{1}{2}\eta_{a,b}\, \frac{\theta^4 \big[ {\textstyle{a\atop b}}\big] (0) }{\eta^{12} }
\\
\tilde{\mathscr{A}} (n\not=0 ) &=\frac{i}{2 \sin ( \pi qR n)} \sum_{a,b}\tfrac{1}{2}\eta_{a,b}\, \frac{\theta \big[ {\textstyle{ a\atop b}}\big] (iq R  n \ell  )\, \theta^3 \big[ {\textstyle{a\atop b}}\big] (0 )}{\vartheta_1 (i qR n \ell  )\, \eta^9 }\,. 
\end{split}
\end{equation}
describing the tree level propagation of closed strings  between the D$p$-branes. From this expression we notice that the massless dilaton, metric and RR forms have the same disk couplings as in the flat space case, while the closed-string Landau levels --- {\em i.e.} those states carrying a non-vanishing winding number --- have the non-trivial couplings
\begin{equation}
\gamma_n = \sqrt{\frac{\text{sgn} (q R n)}{2 \sin (\pi q R n)}} = \frac{1}{\sqrt{\left|2 \,\sin (\pi q R n )\right|}}\,.
\end{equation}
In the field-theory limit, {\em i.e.} for weak magnetic field, one finds 
\begin{equation}
\gamma_{n} \to \frac{1}{\sqrt{|2 \pi q R n|}}\,, \label{onepointst}
\end{equation}
which indeed reproduces the coupling of the dilaton to the Melvin background in the effective low-energy Dirac-Born-Infeld action \cite{Dudas:2001ux}.

\section{Open Strings on $\text{Melvin}\times T^2 \times K3$}

We can finally move to describe systems of D-branes on the anti-self-dual Melvin background with reduced supersymmetry, so that we can compare with the topological amplitudes for open strings \cite{Antoniadis:2013epe}. To this end we consider a set of $M$ D5-branes on the tip of the $\mathbb{C}^2 /\mathbb{Z}_N$ singularity, where the $\mathbb{Z}_N$ acts as opposite-angle rotations on the two $\mathbb{C}$ planes. The Melvin deformation is then along the D-branes world-volume and we assume that the twist is fibered on the $a$-cycle of the $T^2$. The $b$-cycle, which without loss of generality we take to be orthogonal to $a$, is a spectator. Of course, it is possible to study the compact $K3$ version, but in this case one should construct proper orientifold vacua, involving orientifold planes and a suitable number and types of D-branes to cancel the RR tadpoles \cite{Angelantonj:2002ct}. The orientifold projection of strings on Melvin space may be rather unconventional and involves orientifold planes at angles \cite{Angelantonj:2002id}, but for
the purpose of our discussion it is sufficient to consider the simpler non-compact configuration without O-planes. 

To regularise the theory we also introduce Wilson lines $a_i$ along the spectator $b$-cycle of radius $\rho$, so that the annulus partition function  reads
 \begin{equation}
 \mathscr{A} =  \frac{2\pi R_1}{N}\, \sum_{i,j=1}^M \sum_{g =0}^{N-1} \int_0^\infty \frac{dt}{t^{3/2}} \, \sum_{\tilde k_1 \in \mathbb{Z}} \sum_{k_2 \in\mathbb{Z}} e^{-\pi (R_1 \tilde k_1)^2 / \alpha ' t}\, e^{-\pi t \left( \frac{k_2}{R_2} + a_i - a_j\right)^2} \, \mathscr{A}_g (\tilde k_1 ) \,,
 \end{equation}
 with 
 \begin{equation}
 \begin{split}
 \mathscr{A}_g (0) &=  - (2 \sin (\pi g /N))^2\, \sum_{a,b} \tfrac{1}{2} \eta_{a,b} \frac{\theta^2 \big[{\textstyle{a\atop b}}\big] \, \theta \big[ {\textstyle{a \atop b}}\big] (g /N )\, \theta \big[ {\textstyle{a \atop b}}\big] (- g /N)}{\eta^6 \, \vartheta_1 (g /N  ) \vartheta_1 (- g /N )  }\,,
 \\
 \mathscr{A}_g (\tilde k_1 \neq 0) &= - \frac{(2 \sin (\pi g /N))^2}{\left[ 2 \sin (\pi q R \tilde k_1) \right]^2} 
 \\
 &\qquad \times \sum_{a,b}\tfrac{1}{2}\eta_{a,b} \, \frac{ \theta \big[ {\textstyle{a \atop b}}\big] (q R_1 \tilde k_1) \, \theta \big[ {\textstyle{a \atop b}}\big] ( - q R_1 \tilde k_1) \, \theta \big[ {\textstyle{a \atop b}}\big] (g/N) \, \theta \big[ {\textstyle{a \atop b}}\big] (-g/N) }{\vartheta_1  (q R_1 \tilde k_1 )\, \vartheta_1  (- q R_1 \tilde k_1 ) \, \vartheta_1 (g /N  ) \vartheta_1 (- g /N  )} \,.
 \end{split} \label{annMK3}
 \end{equation}
Actually, the previous expression simplifies drastically if one uses the Jacobi identity \eqref{RiemannID}
\begin{equation}
\mathscr{A} = 2 \pi R_1 \,  \sum_{i,j=1}^M \int_0^\infty \frac{dt}{t^{3/2}}\, \sum_{\tilde k_1 \neq 0} \sum_{k_2\in\mathbb{Z}}\, \frac{e^{-\pi (R_1\tilde k_1)^2/\alpha ' t - \pi t \left(\frac{k_2}{R_2} + a_i - a_j \right)^2}}{\left[ 2 \sin (\pi  q R_1 \tilde k_1) \right]^2} \,,
\end{equation}
and reproduces the open-string topological amplitudes of \cite{Antoniadis:2013epe}.  Notice that the tower of string excitations decouples and, as expected, only the open-string BPS states contribute, since the higher-derivative F-terms are BPS protected. 

The integral can now be computed and, in the $a_i - a_j \to 0$ limit where gauge symmetry enhancement occurs,  yields
\begin{equation}
\mathscr{A} \propto  \sum_{i,j=1}^M \sum_{n =1}^\infty \frac{1}{ n }\, \frac{e^{-2\pi R_1 n |a_i - a_j|}}{\left[ \sin (\pi q R_1 n) \right]^2} = 4 \sum_{i,j=1}^M \sum_{n=1}^\infty \frac{1}{n} \, \frac{e^{-2\pi R_1 n |a_i -a_j|}}{(e^{2 i \pi q R_1  n} -1 ) ( 1 - e^{-2 i \pi q R_1  n})}\,. 
\end{equation}
which is precisely the result of Nekrasov \cite{Nekrasov:2003rj} for the $\beta$-deformed $\varOmega$ background with $\epsilon_1 = - \epsilon_2$, once we identify the radius $R_1$ with $\beta$ and $q$ with the single $\epsilon$ parameter.

\subsection{The $\mathscr{N}=2^*$ case}

Also in this case, it is rather straightforward to generalise the previous  construction to the case where an $\mathscr{N}=2^*$ gauge theory is supported on the $M$ D-branes. Following \cite{Florakis:2015ied,Angelantonj:2017qeh, Samsonyan:2017xdi}, a mass for the adjoint hypermultiplet can be generated by making the $K3$ orbifold freely acting, whereby the $\mathbb{Z}_N$  projection on the $\mathbb{C}^2$ space is accompanied by an order-$N$ shift along the spectator circle. As a result, the adjoint hypermultiplet is no longer projected out from the spectrum but gets a mass inversely proportional to the radius of the $S^1$. 

This freely-acting construction requires a minimal modification of the annulus amplitude which, in the absence of Wilson lines, reads
 \begin{equation}
 \begin{split}
 \mathscr{A} &=  \frac{2\pi R_1}{N} \sum_{g =0}^{N-1} \int_0^\infty \frac{dt}{t^{3/2}} \, \sum_{\tilde k_1 \in \mathbb{Z}} \sum_{k_2\in\mathbb{Z}} e^{-\pi (R_1 \tilde k_1)^2 / \alpha ' t}\, e^{-\pi t (k_2/R_2)^2} \, e^{2 i\pi k_2 g/N}\, \mathscr{A}_g (\tilde k_1 ) 
 \\
 &= 2 \pi R_1 \,  \sum_{g=0}^{N-1} \int_0^\infty \frac{dt}{t^{3/2}}\, \sum_{\tilde k_1 \neq 0} \sum_{k_2\in\mathbb{Z}}\, \frac{\left[ 2 \sin (\pi g /N) \right]^2}{\left[ 2 \sin (\pi  q R_1 \tilde k_1) \right]^2}\,   e^{-\pi (R_1\tilde k_1)^2/\alpha ' t - \pi t (k_2/R_2 )^2}\, e^{2 i\pi  k_2 g/N}
 \\
 &\equiv \sum_{k_2\in \mathbb{Z}} \sum_{g=0}^{N-1} \frac{\left[ 2 \sin (\pi g /N) \right]^2\, e^{2 i\pi  k_2 g/N}}{N}\, \mathscr{B} (k_2 ) 
 \end{split}
 \end{equation}
with the $\mathscr{A}_g (\tilde k_1 )$ as in eqs. \eqref{annMK3}, while $\mathscr{B} (k_2 )$ is unambiguously defined by the last identity above. As usual, in writing the second equation we have made use of the Jacobi identity \eqref{RiemannID}. Following \cite{Angelantonj:2017qeh}, the sum over $g$ constrains the allowed Kaluza-Klein momentum $k_2$,
\begin{equation}
\sum_{k_2\in \mathbb{Z}} \sum_{g=0}^{N-1} \frac{\left[ 2 \sin (\pi g /N) \right]^2\, e^{2 i\pi  k_2 g/N}}{N}\,  \mathscr{B} (k_2 ) = \left( 2\, \sum_{k_2 =0 \, \text{mod} \, N} - \sum_{k_2 = \pm 1\, \text{mod} \, N} \right) \, \mathscr{B} (k_2 ) \,,
\end{equation}
where the first sum in the RHS refers to the massless $\mathscr{N}=2$ gauge multiplet while the second one is ascribed to the massive hypermultiplet. 

In the limit where the Kaluza-Klein excitations $k_2$ can be neglected, {\em i.e.} in the limit of large $N$ so that the hypermultiplet mass, $m_\text{H} = 1/R_2$, is much smaller than the other Kaluza-Klein masses, the integral can be straightforwardly computed and yields the expected (field theory) result.

\section{Comments on Melvin and  $\varOmega$ backgrounds}

The fact that the free energies of heterotic and open strings on the Melvin background reproduce the topological amplitudes and, in the field theory limit, the Ne\-krasov free energy should not be a surprise. After all, the $\varOmega$ deformation of maximally supersymmetric or $\mathscr{N}=2$ gauge theories can itself be defined by starting with a gauge theory on a six manifold $\mathbb{R}^4 \times T^2$ \cite{Nekrasov:2003rj}, where the $T^2$ reduction involves an $\mathbb{R}^4$ bundle with non-trivial flat $SO(4)$ connections. In the notation of  \cite{Nekrasov:2003rj}, the resulting metric reads
\begin{equation}
ds^2 = A \, dz \, d\bar z + g_{IJ} ( dx^I + V^I dz + \bar V^I d\bar z ) ( dx^J+ V^J dz + \bar V^J d\bar z )\,, \label{NOk}
\end{equation}
with $A$ the area of the $T^2$, and $V^I = \varOmega^I{}_J x^J$, $\bar V^I = \bar\varOmega^I{}_J x^J$. If $\varOmega^I{}_J$ is anti-self-dual  a residual supersymmetry is preserved.  In this case the geometry \eqref{NOk} is nothing but the Melvin space of eq.~\eqref{gen2Melvin} and, therefore, we have computed the string uplift of the (perturbative) Nekrasov free-energy in the case of a single parameter, $\epsilon_1 = -\epsilon_2$. 

A possible relation between the $\varOmega$ background and the NS fluxbrane solution --- which is nothing but a synonymous for the Melvin solution --- has been already discussed in \cite{Hellerman:2011mv, Hellerman:2012zf, Orlando:2013yea, Lambert:2014fma}. In these papers, the authors  study the dynamics of D-branes in the NS fluxbrane background and, upon a T-duality along the compact coordinate, they notice that the derivative expansion of the Dirac-Born-Infled action on this background actually reproduces the $\varOmega$-deformed super-Yang-Mills theory. As a result, the authors made the correct conclusion that the Melvin background is indeed the string theoretic realisation of the $\varOmega$ deformation of flat space-time in the case $\epsilon_1 = -\epsilon_2$.

In the D3-brane realisation of the low-energy gauge theory, one may perform an S duality on the parent type IIB superstring. To study the effect of this non-perturbative duality, it is simpler to start from the T-dualised Melvin background, whereby the off-diagonal components of the metric are mapped into a non trivial background for the Kalb-Ramond field. Therefore, the S duality  transforms it into a non-trivial RR two-form
\begin{equation}
B_2 \to C_2\,,
\end{equation}
while the metric gets Weyl transformed and the dilaton flips sign. It was suggested in \cite{Billo:2006jm,Billo:2009di} that this RR background might reproduce the $\varOmega$ deformation of Nekrasov. However, although this seems to be true at the linear order in $\epsilon$, the D3-brane dynamics in the RR fluxtrap solution is not governed by the equivariant action of Nekrasov and Okounkov \cite{Nekrasov:2003rj}. 

The correct identification of $\varOmega$-background with the Melvin space \eqref{ASDMelvin} allows for a straightforward generalisation to the case of independent 
rotations, $\epsilon_1 \neq \epsilon_2$, and we hope to come back to this point in the near future.

\section*{Acknowledgements} We thank Sergio Ferrara, Ioannis Florakis, Elias Kiritsis, Kumar Narain, Domenico Orlando, Susanne Reffert and Domenico Seminara for stimulating discussions.  This work was supported in part by the Swiss National Science Foundation, in part by the Labex ``Institut Lagrange de Paris''  in part by a CNRS PICS grant and in part by the MIUR-PRIN contract 2017CC72MK\_003. C.A. would like to thank the Albert Einstein Center for Fundamental Physics of the University of Bern, the Theory Division of CERN, the CPHT of Ecole Polytechnique and the LPTHE of Sorbonne University for hospitality during various stages of this work. We also thank the Galileo Galilei Institute for Theoretical Physics and INFN for hospitality and partial support during the workshop ``String Theory from a worldsheet perspective'' where part of this work has been done.

\appendix\section{Dedekind Eta Function and Jacobi Theta Functions}\label{appJacobi}

In this Appendix we collect our conventions for the Dedekind $\eta$ and Jacobi theta functions. To start the {\em nome} $q$ is related to the complex structure of the world-sheet torus via $q=e^{2 i \pi\tau}$. The Dedekind function is defined by the infinite product
\begin{equation}
\eta (q) = q^{\frac{1}{24}}\, \prod_{n=1}^\infty (1 - q^n )\,,
\end{equation}
and transforms as follows under the action of the generators of the modular group $\text{SL} (2;\mathbb{Z} )$,
\begin{equation}
\eta (\tau+1) = e^{\frac{i\pi}{12}}\, \eta (\tau )\,, \qquad
\eta (-1/\tau ) = (-i\tau )^{1/2}\, \eta (\tau )\,.
\end{equation}
Theta functions with generic characteristics $a$ and $b$ and non-trivial argument $z$ admit the following representations in terms of infinite sum
\begin{equation}
\theta {\textstyle \big[ {a\atop b}\big]} (z |\tau ) = \sum_{n\in \mathbb{Z} } q^{\frac{1}{2} (n+a)^2}\, e^{2 i \pi (n+a)(b+z)}\,,
\end{equation}
and in terms of infinite product
\begin{equation}
\begin{split}
\theta {\textstyle \big[ {a\atop b}\big]} (z |\tau ) &= e^{2 i \pi a (b+z)} \, q^{\frac{1}{2} a^2}\, \prod_{n=1}^\infty (1-q^n ) 
\\
&\qquad \qquad \times \left( 1 + q^{n+a-\frac{1}{2}}\, e^{2 i \pi (b+z)} \right) \left( 1+ q^{n-a-\frac{1}{2}} \,e^{-2i\pi (b+z)}\right)\,.
\end{split}
\end{equation} 
Whenever $a$ and $b$ take the values $0, \frac{1}{2}$, they reduce to the four functions
\begin{equation}
\begin{split}
\vartheta_1 (z|\tau ) &= - \theta {\textstyle \big[{1/2\atop 1/2}\big]} (z|\tau )\,,
\\
\vartheta_3 (z|\tau ) &= \theta {\textstyle \big[{0\atop 0}\big]} (z|\tau )\,, 
\end{split}
\qquad 
\begin{split}
\vartheta_2 (z|\tau ) &= \theta {\textstyle \big[{1/2 \atop 0}\big]} (z|\tau )\,,
\\
\vartheta_4 (z|\tau ) &= \theta {\textstyle \big[{0\atop 1/2}\big]} (z|\tau )\,.
\end{split}
\end{equation}
The theta functions have a simple behaviour with respect to the modular transformations,
\begin{equation}
\begin{split}
\theta {\textstyle \big[ {a\atop b}\big]} (z|\tau + 1 ) &= e^{-i \pi a (a-1)}\, \theta {\textstyle \big[ {a\atop b+a-1/2}\big]} (z|\tau ) \,,
\\
\theta {\textstyle \big[ {a\atop b}\big]} (z/\tau |-1/\tau ) &= (-i \tau)^{1/2} \, e^{2i\pi ab + i\pi z^2/\tau}\, \theta {\textstyle \big[ {b\atop -a}\big]} (z |\tau )\,.
\end{split}
\end{equation}

The Jacobi theta functions obey a series of non-trivial identities. Those of interest to us are
\begin{equation}
\begin{split}
\sum_{a,b} \eta_{a,b} \, \prod_{i=1}^4 \theta {\textstyle\big[{a\atop b}\big]} (z_i |\tau ) &= - 2 \prod_{i=1}^4 \vartheta_1 (\tilde z_i |\tau )\,,
\\
\sum_{a,b} \hat\eta_{a,b} \, \prod_{i=1}^4 \theta {\textstyle\big[{a\atop b}\big]} (z_i |\tau ) &= + 2 \prod_{i=1}^4 \vartheta_1  (\hat z_i |\tau )\,,
\end{split}\label{RiemannID}
\end{equation}
where $\eta_{a,b} = (-1)^{2a + 2b + 4ab}$ and $\hat \eta_{a,b} = (-1)^{2a+2b}$ denote the GSO phases for the type IIB and type IIA superstrings, respectively, and
\begin{equation}
\begin{split}
\tilde z_1 &= \tfrac{1}{2} (z_1 + z_2 + z_3 - z_4 )\,,
\\
\tilde z_2 &= \tfrac{1}{2} (z_1 + z_2 - z_3 + z_4 )\,,
\\
\tilde z_3 &= \tfrac{1}{2} (z_1 - z_2 + z_3 + z_4 )\,,
\\
\tilde z_4 &= \tfrac{1}{2} (-z_1 + z_2 + z_3 + z_4 )\,,
\end{split}
\qquad
\begin{split}
\hat z_1 &= \tfrac{1}{2} (z_1 + z_2 + z_3 + z_4 )\,,
\\
\hat z_2 &= \tfrac{1}{2} (z_1 + z_2 - z_3 - z_4 )\,,
\\
\hat z_3 &= \tfrac{1}{2} (z_1 - z_2 + z_3 - z_4 )\,,
\\
\hat z_4 &= \tfrac{1}{2} (z_1 - z_2 - z_3 + z_4 )\,.
\end{split}
\end{equation}
Whenever one or more of the combinations $\tilde z_i$ or $\hat z_i$ equal zero, the right-hand side vanishes. Typically, the number of vanishing $\tilde z_i$, or $\hat z_i$ counts the number of supercharges which are preserved by the background. This is true, for instance, in four and six dimensions, but not in $d=2$ where supersymmetry does not imply a precise degeneracy between fermionic and bosonic states \cite{Hull:1985jv}.

Finally, the theta functions satisfy the following periodicity relations
\begin{equation}
\begin{split}
\theta {\textstyle\big[ {a +m \atop b+n}\big]} (z|\tau ) &=  e^{2 i \pi a n}\, \theta {\textstyle\big[ {a \atop b}\big]} (z|\tau )\,,
\\
\theta {\textstyle\big[ {a \atop b}\big]} (z +m \tau + n|\tau ) &= e^{2i\pi (a n - m (b+z))}\, q^{-m^2/2}\,  \theta {\textstyle\big[ {a \atop b}\big]} (z |\tau )
 \,.
 \end{split}
\label{JThPeriodicity}
\end{equation}

A useful identity is 
\begin{equation}
\begin{split}
\left( \frac{2 \pi i  \lambda \, \eta^3}{\vartheta_1 (\lambda |\tau )}\right)^2\, e^{-\pi \lambda^2 /\tau_2} &= - \exp \left\{ \sum_{k=1}^\infty 
\frac{1}{k} \hat G_{2k} (\tau ) \lambda^{2k} \right\}
\\
&= - \sum_{k=0}^\infty S_{2k} \left(\hat G_2 , \frac{1}{2} G_4 , \ldots , \frac{1}{k} G_{2k} \right)\, \lambda^{2k} \,,
\end{split} \label{theta1id}
\end{equation}
with 
\begin{equation}
G_{2k} (\tau ) \equiv  2 \zeta (2k)\, E_{2k} (\tau) =  2 \zeta (2k)\, \left( 1 + \frac{(2\pi i )^{2k}}{(2k-1)!\, \zeta (2k)} \, \sum_{n=1}^\infty \sigma_{2k-1} (n) \, q^n\right)\,,
\end{equation}
the holomorphic Eisenstein series.\footnote{Note that $G_2$ is not modular invariant and thus we have to replace it by $\hat G_2 = G_2 - \frac{\pi}{\tau_2}$.}

In this expression $S_{2k}$ are the Schur polynomials defined via the identity
\begin{equation}
\sum_{k=0}^\infty S_k (x_1 , \ldots , x_k )\, z^k = \exp\, \sum_{k=1}^\infty x_k \, z^k\,.
\end{equation}

\section{Some properties of the two-dimensional Narain lattice}\label{NarainlattB}

In this appendix we list some properties of the two-dimensional Narain lattice, which are essential to match the expression of the vacuum energy of the heterotic string on the Melvin background and the heterotic topological amplitudes. 

We first recall the definition of the complex momenta
\begin{equation}
\begin{split}
p_\text{L} &= \frac{1}{\sqrt{T_2 U_2}} \left( m_2 - U m_1 + \bar T ( n_1 + U\, n_2 )\right)\,,
\\
p_\text{R} &= \frac{1}{\sqrt{T_2 U_2}} \left( m_2 - U m_1 +  T ( n_1 + U\, n_2 )\right)\,,
\end{split} \label{Complexp}
\end{equation}
written in terms of the complex structure and K\"ahler moduli $U$ and $T$, and the quantised momenta $m_{1,2}$ and winding numbers $n_{1,2}$. The Narain lattice in the Hamiltonian representation reads
\begin{equation}
\Lambda (T,U) = \sum_{m_i,n_i} q^{\frac{1}{4} |p_\text{L}|^2} \, \bar q ^{\frac{1}{4} |p_\text{R} |^2 }\,,
\end{equation}
Upon a Poisson summation over the Kaluza-Klein momenta, we can cast the previous expression in the Lagrangian representation
\begin{equation}
\begin{split}
\Lambda (T,U) &= \frac{T_2}{\tau_2} \sum_{\tilde m_i , n_i} e^{2 \pi i T \det (A)}\, \exp \left\{ - \frac{\pi T_2}{\tau_2 U_2} \, \left| \begin{pmatrix} 1 & \tau \end{pmatrix} \, A\, \begin{pmatrix} 1 \\ U \end{pmatrix}\right|^2 \right\}
\\
&= \frac{T_2}{\tau_2} \sum_{\tilde m_i , n_i} \Lambda_{\tilde m_i , n_i} (T, U)\,.
\end{split}\label{NarainL}
\end{equation}
The two-by-two matrix $A$ encodes the winding numbers
\begin{equation}
A = \begin{pmatrix} \tilde m_1 & \tilde m_2 \\ n_1 & n_2 \end{pmatrix} \,.
\end{equation}
The two-dimensional Narain lattice also obeys the remarkable relation
\begin{equation}
\sum_{m_i , n_i} \left( \frac{\tau_2 p_\text{L}}{\sqrt{T_2 U_2}}\right)^N \, q^{\frac{1}{4} |p_\text{L}|^2} \, \bar q ^{\frac{1}{4} |p_\text{R} |^2 } = \frac{T_2}{\tau_2} \sum_{\tilde m_i , n_i} \left( \frac{ \tilde m_1 +  U \tilde m_2 + \bar \tau (n_1 +  U n_2 )}{ U_2}\right)^N\, \Lambda_{\tilde m_i , n_i} (T, U)\,.
\end{equation}

Combining this result with the expansion \eqref{theta1id} we can derive the useful identity
\begin{equation}
\sum_{m_i, n_i} \left( \frac{2 \pi i  \lambda \, \eta^3}{\vartheta_1 (\tilde\lambda |\tau )}\right)^2\, e^{-\pi \tilde\lambda^2 /\tau_2} \, q^{\frac{1}{4} |p_\text{L}|^2}\, \bar q ^{\frac{1}{4} |p_\text{R}|^2} = \frac{T_2}{\tau_2}
\sum_{\tilde m_i , n_i} \left( \frac{2 \pi i  \lambda \, \eta^3}{\vartheta_1 (\hat\lambda |\tau )}\right)^2\, e^{-\pi \hat\lambda^2 /\tau_2} \, \Lambda_{\tilde m_i , n_i } (T,U)
\label{theta1id}
\end{equation}
where
\begin{equation}
\tilde\lambda = \frac{\lambda\tau_2 p_\text{L}}{\sqrt{T_2 U_2 }}\,, \qquad \hat\lambda = \frac{\lambda \hat p_\text{L}}{U_2}\,,
\label{defpth}
\end{equation}
with $\hat p_\text{L} = \tilde m_1 + U \tilde m_2 + \bar \tau (n_1 + U n_2 )$.

\bibliographystyle{unsrt}

\end{document}